\documentclass[namedreferences]{SolarPhysics}
\usepackage[optionalrh]{spr-sola-addons} 
\usepackage{graphicx}                    
\usepackage{color}                       
\usepackage{url}                         

\newcommand{\aap}{    {\it Astron. Astrophys.}}

\newcommand{\apj}{    {\it Astrophys. J.}}

\newcommand{\mnras}{  {\it Mon. Not. Roy. Astron. Soc.}}

\newcommand{\pasj}{   {\it Pub. Astron. Soc. Japan}}

\newcommand{\solphys}{{\it Solar Phys.}}

\def\gca{Geochim.~Cosmochim.~Acta}%

\def\kms{$\mathrm{km\, s^{-1}}$}
\newcommand{\COBOLD}{{\sf CO$^5$BOLD}}
\newcommand{\cobold}{\COBOLD}

\newcommand{\xx}{\ensuremath{\mathrm{1D}_{\mathrm{LHD}}}}
\newcommand{\mD}{\ensuremath{\left\langle\mathrm{3D}\right\rangle}}
\newcommand{\loggf}{\ensuremath{\log\,gf}}
\DeclareRobustCommand{\ion}[2]{%
\relax\ifmmode
\ifx\testbx\f@series
{\mathbf{#1\,\mathsc{#2}}}\else
{\mathrm{#1\,\mathsc{#2}}}\fi
\else\textup{#1\,{\mdseries\textsc{#2}}}%
\fi}

\begin{document}

\begin{article}

\begin{opening}

\title{Solar Chemical Abundances Determined with a CO5BOLD 
3D Model Atmosphere}

%
\author{E.~\surname{Caffau}$^{1}$\sep
        H.-G.~\surname{Ludwig}$^{1,2,3}$\sep
        M.~\surname{Steffen}$^{4}$\sep
        B.~\surname{Freytag}$^{5}$\sep
        P.~\surname{Bonifacio}$^{1,2,6}$
       }

%
\runningauthor{E.~Caffau et al.}
\runningtitle{Solar Abundances}

%
\institute{$^{1}$ GEPI, Observatoire de Paris, CNRS, Universit\'e Paris Diderot; 92195
Meudon Cedex, France \\
$^{2}$ CIFIST Marie Curie Excellence Team [\url{http://cifist.obspm.fr}] \\
$^{3}$ Zentrum f\"ur Astronomie der Universit\"at Heidelberg, Landessternwarte, 
K\"onigstuhl 12, 69117 Heidelberg, Germany \\
$^{4}$ Astrophysikalisches Institut Potsdam, An der Sternwarte 16, D-14482 Potsdam, Germany \\
$^{5}$ CRAL,UMR 5574: CNRS, Universit\'e de Lyon,
\'Ecole Normale Sup\'erieure de Lyon,
46 all\'ee d'Italie, F-69364 Lyon Cedex 7, France \\
$^{6}$ Istituto Nazionale di Astrofisica,
Osservatorio Astronomico di Trieste,  Via G.B. Tiepolo 11,
I-34143 Trieste, Italy
}

\begin{abstract}
In the last decade, the photospheric solar metallicity as 
determined from spectroscopy experienced a remarkable downward revision.
Part of this effect can be attributed to an improvement of atomic 
data and the inclusion of NLTE computations, but also the use of 
hydrodynamical model atmospheres seemed to play a role.
This ``decrease'' with time of the metallicity of the solar photosphere
increased the disagreement with the results from helioseismology.
With a CO5BOLD 3D model of the solar atmosphere, the CIFIST team
at the Paris Observatory re-determined the photospheric solar abundances
of several elements, among them C, N, and O.
The spectroscopic abundances are obtained by fitting the
equivalent width and/or the profile of observed spectral lines
with synthetic spectra computed from the 3D model atmosphere.
We conclude that the effects of granular fluctuations depend on 
the characteristics of the individual lines, but are found to be 
relevant only in a few particular cases. 
3D effects are not reponsible for the systematic
lowering of the solar abundances in recent years.
The solar metallicity resulting from this analysis is $Z=0.0153$, $Z/X=0.0209$.
\end{abstract}

%

\end{opening}

%
\section{Introduction}

Since the pioneering work of \inlinecite{russell29}, the study of the chemical
composition of the solar photosphere has been an important topic
in astronomy. Russell, measuring the strength of the absorption lines
in the observed solar spectrum, determined the abundance of 56 elements
and six molecules.
After this work many other astronomers analysed the solar
spectrum in order to deduce the detailed pattern of photospheric abundances.
The accurate determination of the solar chemical abundances is a major 
topic, because:
\begin{enumerate}
\item
the knowledge of the elemental abundances in the present solar photosphere 
is the basis to infer the chemical composition of the initial Sun, and allows one
to reconstruct the past and future evolution of the Sun, including its 
physical and chemical internal structure;
\item
the comparison of the chemical abundances in the solar photosphere and 
in meteoritic samples provides important information about the formation
and chemical evolution of the solar system;
\item
the construction of detailed models of the solar atmosphere requires
the knowledge of its chemical composition;
\item
the knowledge of the solar photospheric chemical abundance
allows the empirical determination of the oscillator strength of 
any spectral line of this element observable in the solar spectrum;
\item
the solar abundances serve as a reference for the chemical analysis 
of other stars in the Galaxy, of the interstellar medium, and of the
stellar populations of external galaxies.
\end{enumerate}
In our analysis of the solar photosphere we are mainly interested 
in the first and third point.

For our work on the solar abundance determinations, we rely on 
a \cobold\ 3D model of the solar photosphere. We were interested to 
understand if the presence of horizontal fluctuations (the solar granulation)
has a systematic effect on the abundances derived from the 3D model
with respect to what is obtained by 1D models which ignore granulation.

We find that the effects of granular fluctuations can lead to both positive
and negative abundance corrections, depending on the properties of the
individual spectral line under consideration. However, the granulation effect
is relevant only in a few particular cases. 
3D effects are not responsible for the systematic
lowering of the solar abundances in recent years.

In the following sections we summarise the investigations of some elements 
we have already published, and present a new analysis of some other elements 
(Li, K, Fe, and Os).
For a complete overview of the solar abundance determinations we recommend
the review by \inlinecite{lodders09} as a very detailed and complete work.

\section{Model Atmospheres}

Our abundance analyses of the solar photosphere are based
on a time-dependent, 3D, hydrodynamical model atmosphere computed
with the \cobold\ code [\url{http://www.astro.uu.se/~bf/co5bold_20020216/cobold.html}]
(\citeauthor{freytag02} \citeyear{freytag02}, 
\citeyear{freytag03}).
The 3D model atmosphere we use has a 
box size of $5.6\times 5.6\times 2.27\,{\rm Mm}^3$, 
a resolution of $140\times 140\times 150$ grid points, and
spans a range in optical depth of about $-6.7<\log\tau_\mathrm{Ross}<5.5$
(from $-1.4$~Mm below to $+0.9$~Mm above $\tau_\mathrm{Ross}=1$).
The selected 19 snapshots we use for the spectrum-synthesis calculations 
are equidistantly spaced in time, sufficiently separated in time to show
little correlation, and cover 1.2\,hours of solar time.
For comparison, the following 1D plane-parallel model atmospheres were
considered:
\begin{enumerate}
\item
the \mD\ model obtained by horizontally averaging each 3D snapshot
over surfaces of equal (Rosseland) optical depth;
\item
the \xx\ model, a 1D hydrostatic mixing-length model which employs the same 
micro-physics and radiative transfer scheme as \cobold;
\item
the semi-empirical Holweger-M\"uller solar model (\citeauthor{hhsunmod}
\citeyear{hhsunmod}, \citeauthor{hmsunmod} \citeyear{hmsunmod}, hereafter HM);
\item
the ATLAS-9 model computed by 
F. Castelli [\url{http://wwwuser.oats.inaf.it/castelli/sun/ap00t5777g44377k1asp.dat}]
with the solar abundances of \inlinecite{sunabboasp}.
\end{enumerate}

The \mD\ and \xx\ models have been introduced as reference 1D models,
because they share the microphysics with the 3D model \cite{zolfito}.
For more details about the 1D and 3D models used in our analysis see 
\inlinecite{oxy} or the other papers of the collaboration on the 
determination of the solar photospheric abundances.

\section{3D Corrections}

We investigated the effects of 3D convection on solar abundances in
\inlinecite{caffau09b}.
To define 3D corrections we selected two reference models: \mD\ and \xx.
Both share the microphysics of the 3D-\cobold\ model to ensure differential
comparability. We distinguish two different 3D corrections, defined as:
${\Delta^{(1)}(\xi_{\rm mic})=A(Y)_{\rm 3D}}-A(Y)_{\rm \mD}$ and
${\Delta^{(2)}(\xi_{\rm mic},\alpha_{\rm MLT})=
A(Y)_{\rm 3D}}-A(Y)_{\rm \xx}$, where $Y$ represents any chemical
element, and $A(Y)=\log{(n_Y/n_{\rm H})}+12$.
The first correction isolates the granulation effects, {\it i.e.} the influence of 
the horizontal fluctuations around the mean stratification, while the second 
one measures the total 3D effect, accounting for both horizontal fluctuations
and for the different influence of 3D hydrodynamical and 1D mixing-length 
convection, respectively, on the resulting \emph{mean} temperature structure.
Obviously, both 3D corrections depend on the choice of the microturbulence 
parameter, [$\xi_{\rm mic}$], used with the 1D models, while the second one 
is also a function of the mixing length parameter, [$\alpha_{\rm MLT}$], adopted 
for the \xx\ model.
The idea is that these 3D corrections can be derived from the simulations
with much better accuracy than the detailed thermal structure of the solar
atmosphere in absolute terms. While the latter depends sensitively on {\it e.g.},
the frequency binning adopted for the radiation hydrodynamics, the 3D 
corrections are less sensitive to such details, as they are defined in a 
strictly differential way. Hence, the 3D corrections are not merely evaluated 
to measure the abundance difference between a given pair of 3D and 1D 
models, but they may rather be utilised to improve the abundance 
determinations from standard 1D models.
Abundances derived from the semi-empirical HM model should
be corrected by adding only the ``granulation correction'' 
$\Delta^{(1)}(\xi_{\rm mic})$, 
while abundances based on standard 1D mixing-length models like ATLAS or 
MARCS, must be corrected by adding the ``total 3D correction'' 
$\Delta^{(2)}(\xi_{\rm mic},\alpha_{\rm MLT})$.

The 3D corrections are well defined as long as the considered lines are
weak such that their equivalent widths are independent of the assumed 
microturbulence. The situation becomes more problematic once the lines
are partly saturated, because then the corrections depend on 
$\xi_{\rm mic}$, and the choice of the microturbulence parameter
is critical. In principle, the microturbulence parameter derived
empirically from analysing observed solar spectra with a standard 1D
model, $\xi_{\rm mic}$(Sun-1X), should be identical with the theoretical
microturbulence $\xi_{\rm mic}$(Hydro-\xx), obtained from analysing
the synthetic spectrum of the 3D hydrodynamical model atmosphere
with the \xx\ model. However, there are indications that 
$\xi_{\rm mic}$(Hydro-\xx) is systematically smaller than 
$\xi_{\rm mic}$(Sun-1X) \cite{mst09}. As a consequence, the 3D corrections 
should be computed with $\xi_{\rm mic}$(Hydro-\xx). Wherever possible, 
abundances from weak lines are to be preferred. 

For some of the weak lines that we investigated ({\it e.g.} the \ion{C}{i} 
line at 538\,nm), we compared the 3D  correction $\Delta^{(1)}$ with 
the results of \inlinecite{mst02}, which are based on a 2D
hydrodynamical simulation, and find a close agreement.
Summarising our analysis, we can say that the total 3D correction 
$\Delta^{(2)}$ is positive for the majority of the relevant spectral 
lines, except for a few weak, high-excitation lines (mostly Nitrogen) 
for which both $\Delta^{(1)}$ and $\Delta^{(2)}$ are negative.
However, the 3D corrections are small in general.

The abundances presented in this work have been derived directly from the
3D model, in line with our previous publications. We trust in the temperature 
structure of our 3D-\cobold\ model, because it is able to reproduce the 
centre-to-limb variation of the continuum intensity even somewhat better 
than the HM model \cite{hgl09}. However, we plan to compute abundances 
also with the alternative approach, {\it i.e.} from the HM model and correcting 
by $\Delta^{(1)}(\xi_{\rm mic})$, in future investigations. While it is not 
entirely clear which of the two approaches gives the more reliable results,
we propose the abundance differences between the two methods to be considered 
as a measure of the systematic uncertainty of the abundance analysis.

\section{Observational Data}

Our analysis is based on mainly four high resolution, high 
signal-to-noise ratio [S/N] spectra, two for disc centre,  and 
two for the integrated disc.
For most elements we use more than one solar spectrum, because we 
realised that the abundance derived from different spectra do not 
always agree within one $\sigma$.
The observed spectra we considered are:

\begin{enumerate}
\item
the integrated disc spectrum based on fifty solar FTS
scans taken by J. Brault and L. Testerman at Kitt Peak between
1981 and 1984 \cite{kuruczflux};
\item
the two absolutely calibrated FTS spectra obtained at 
Kitt Peak in the 1980s, covering the range 330\,nm to 1250\,nm for 
the integrated disc and disc centre \cite{neckelobs,neckel1999};
\item
the disc centre intensity spectrum in the 
range 300\,nm to 1000\,nm observed from the 
Jungfraujoch \cite{delbouille}, and in the range from 1000\,nm to 5400\,nm
observed from Kitt Peak \cite{delbouilleir}.
\end{enumerate}

\section{Chemical Abundances}

\subsection{Lithium}

Lithium is widely studied in metal-poor stars.
The \ion{Li}{i} resonance doublet at 670.7\,nm is also
observable in the solar spectrum and in the spectra of F--K main 
sequence stars. In metal-poor stars, this region is very clean,
and the Lithium feature is not blended. This is not the case in 
solar-metallicity stars, where atomic and molecular lines contaminate 
the region. Good atomic data for these blending lines are not readily
available in databases, but there are a few published line lists that 
have been used for the abundance analysis of the solar photosphere.
For our own analysis of Li in the solar photosphere, we took into account
both granulation and 3D-NLTE\footnote{NLTE stands for Non Local Thermodynamic Equilibrium} 
effects in computing the
contribution of the Li doublet, while 3D-LTE line formation calculations
were performed for the other blend components.

\begin{figure} 
 \centerline{\includegraphics[width=1.0\textwidth,clip=true]{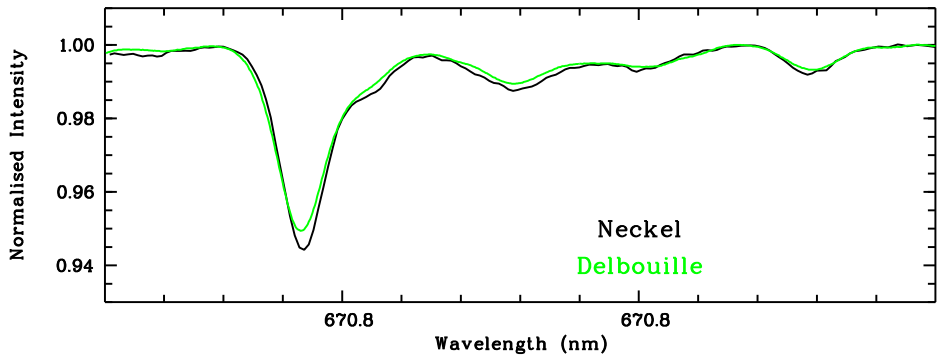}}
 \caption{Comparison of the two atlases of the solar disc centre
 in the region of the \ion{Li}{i} line.
 }
 \label{fig:liint}
 \end{figure}

Considering the four solar atlases described above to investigate lithium,
we realise that the two observed intensity spectra are significantly 
different (see Figure\,\ref{fig:liint}). 
We have no explanation for this disagreement. 
The easiest way to explain it would be to invoke telluric absorption,
but this spectral range has been carefully scrutinised in the
context of the Li isotopic ratio in metal-poor stars, and it seems
unlikely that a telluric absorption could have gone unnoticed.
We decided to discard the Delbouille disc centre atlas, and
to work only with the other three atlases for the Li abundance determination.
The reason for discarding the Delbouille disc centre atlas is that we cannot
reproduce the profile with synthetic spectra, while we can in the
case of the other three solar atlases, and obtain Li abundances
that agree closely.

The two lists of blending lines that can reproduce the solar spectrum
in this region reasonably well are those from \inlinecite{ghezzi09} 
and from \inlinecite{reddy02}.
To obtain the Li abundance, we fit the observed line profile,
interpolating in a grid of 3D synthetic spectra, in which
NLTE effects are taken into account for Li.
We obtain $A$(Li)$_{\rm 3D-NLTE}=1.03\pm 0.03$, fitting the three
observed line profiles using the \inlinecite{ghezzi09} line list.
Fortunately, the solar Lithium abundance is not very sensitive to 
the choice between the two line lists: when using the \inlinecite{reddy02}
list, we obtain $A$(Li)$_{\rm 3D-NLTE}=1.01$.

\subsection{Carbon}

For a selection of 45 lines, we measured the equivalent widths, taking into 
account the blending components, when present, and we obtained 
$A$(C)$=8.50\pm 0.06$ \cite{carbon}. NLTE corrections were computed with the
Kiel code \cite{SH84}, using the \mD\ model as input.
In our sample of \ion{C}{i} lines there is one forbidden line at 872.7\,nm, 
formed in LTE conditions. Its abundance
of A(C)=$8.48\pm 0.02$ is very close to the value we obtain from the complete sample.
A considerable fraction of our selected lines are strong, in the saturated part of the 
curve of growth. We do not see a trend of the Carbon abundance with respect to
the equivalent widths of the lines, and the A(C) is unchanged if we
consider only lines whose EW is smaller than 12\,pm. When we consider lines
weaker than 8\,pm of equivalent width, A(C) is lower of 0.02\,dex.

\subsection{Nitrogen}

Our analysis relies on two sets of equivalent width measurements 
\cite{biemont90,grevesse90}. The Nitrogen abundance is obtained by
interpolating in synthetic curves-of-growth from 3D and 1D models.
To account for NLTE effects, we applied the corrections computed in 
using the \mD\ as the 1D background stratification.
The result is $A$(N)$=7.86\pm 0.12$, as reported in detail by
\inlinecite{azoto} and \inlinecite{maiorca09}.
\inlinecite{asplund09} criticise our line selection, but unfortunately 
provide no information on their own analysis, except their line-to-line 
scatter of 0.04\,dex.
We note that the difference between the abundance from their atomic and
molecular NH $\Delta v=1$ lines is about 0.1\,dex.

\subsection{Oxygen}

Our photospheric Oxygen abundance of $8.76\pm 0.07$ is derived from ten
atomic Oxygen lines. The NLTE corrections are computed from the
\mD\ model as input to the Kiel code, and applied to the 3D-LTE abundances.
For details of the analysis see \inlinecite{oxy} and \inlinecite{oxysanya}.
In our analysis we find a disagreement between the two forbidden [OI] lines
at 630\,nm and 636\,nm of more than 0.1\,dex that we are not able to explain, 
unless we allow the contribution of the Nickel blend of the line at 630\,nm 
to be smaller than expected by about a factor of two (see also 
\opencite{ayres08}). Interestingly, \inlinecite{lambert78a} already
considered the presence of the \ion{Ni}{i} line, but concluded that such 
contribution must be small to have an agreement between the two [OI] lines.
This disagreement is not present in \inlinecite{asplund04},
while in \inlinecite{asplund09} a disagreement is recognised, but 
no solution is offered. When we use the equivalent widths given by
\inlinecite{asplund04} together with our 3D model, we find indeed an 
agreement between the two lines, but with a ``high'' $A$(O) value of 
8.79 and 8.76, respectively, while \inlinecite{asplund04} give 
$A$(O) = 8.69 and 8.67 for the two forbidden lines. This difference 
is probably related to differences in the temperature structure of 
the two independent 3D model atmospheres.

\subsection{Phosphorus}

We considered five infrared \ion{P}{i} lines of Multiplet\,1 for the
photospheric abundance determination \cite{phosphorus}.
Our photospheric Phosphorus abundance of $A$(P)$=5.46\pm 0.04$ is 
in perfect agreement with the meteoritic value \cite{lodders09}.
The Phosphorus abundance in \inlinecite{asplund09}
is 0.06\,dex larger with respect to the value of $A$(P)$=5.35\pm0.04$ 
given in \inlinecite{sunabboasp}. It is unclear whether this upward
revision is related to their new 3D model or to a change in the line list,
in the \loggf\, values, and/or the adopted equivalent widths.

\subsection{Sulphur}

In \inlinecite{lodders09} the photospheric Sulphur abundance is given
as $A$(S)$=7.14\pm 0.01$ and the meteoritic one as $A$(S)$=7.17\pm 0.02$.
We have studied several \ion{S}{i} lines in the solar spectrum.
From the weak forbidden line at 1082\,nm we obtain
$A$(S)$=7.15\pm\left(0.01\right)_{\rm stat}\pm\left(0.05\right)_{\rm sys}$ 
\cite{zolfito}. \inlinecite{asplund09} criticise both the measured 
equivalent widths as being too small, and the \loggf\ we (and also
\inlinecite{ryde06}) use as being obsolete, claiming that updated
values should be used. As no further details are given, it is 
impossible to judge whether their equivalent widths and \loggf\, values
are better. We considered also the permitted lines of Multiplet\,3, Multiplet\,6, 
and Multiplet\,8, discarding the strong lines of Multiplet\,1 at 920\,nm, because
they are blended with telluric absorption \cite{zolfo}. Both lines of 
Multiplet\,6 and Multiplet\,8 are weak and close to LTE. The abundance we find 
is $A$(S)$=7.14$ from the line of Multiplet\,8, and $A$(S)$=7.11$ from the two 
lines of Multiplet\,6. The lines of Multiplet\,3 at 1045\,nm are affected by 
departures from LTE. If we take this effect into account, using the NLTE 
correction of \inlinecite{takeda05}, we obtain $A$(S)$=7.30$. This value 
becomes $A$(S)$=7.28$ when using the NLTE corrections computed by S. Andrievsky 
and S. Korotin with the Sulphur model atom described in \inlinecite{korotin08} and 
\inlinecite{korotin09}. The simple average over all of the multiplets would give 
$A$(S)$=7.17\pm 0.07$, where $\sigma$ is the root-mean-square deviation.
Giving twice the weight to the [SI] lines and the lines of Multiplet\,8 because they are
unaffected by NLTE, we obtain $A$(S)$=7.16\pm 0.05$, which we recommend as
the solar photospheric value.

\subsection{Potassium}

In the solar spectrum, Potassium is observed through lines of 
the neutral species, \ion{K}{i}.
As for other alkalines, the strongest lines are those of the
resonance doublet of \ion{K}{i} at 766.4\,nm and 769.8\,nm, the only 
observable lines in metal poor stars. 
In fact, the other \ion{K}{i} lines are much weaker.
Unfortunately, the strongest \ion{K}{i} line at 766.4\,nm is 
heavily blended in the solar spectrum by strong telluric O$_2$ absorption.

\inlinecite{delareza75} 
performed the first detailed study of the \ion{K}{i} 769.8\,nm 
doublet component in the solar photosphere, analysing its
centre-to-limb variation and computing the K abundance in NLTE.

Among the few excited \ion{K}{i} lines, the one at 404.4\,nm 
is often used in old solar abundance analyses; but this line is 
significantly blended and the $f$ value is uncertain so that
it is excluded by \inlinecite{delareza75}. Also \inlinecite{lambert78b} 
suggest that a wrong $f$ value could be the reason for the exceedingly
low abundance derived from this line.

\inlinecite{takeda96} obtain a good fit of the solar flux 
spectrum with their synthetic profile of the line at 769.8\,nm with a
multi-parameter fitting method. They also studied the dependence of the
K abundance upon the atomic parameters, the microturbulence, and the adopted 
model atmosphere (see their Table~4). The NLTE correction for this line is
estimated  by them to be about $-0.4$~dex.

\inlinecite{zhang06} consider seven lines: the line at 693.8\,nm, 
the resonance doublet, plus four lines in common with \inlinecite{lambert78b}.
These authors derive the Potassium abundance by line profile fitting of the
integrated disc solar spectrum with a NLTE synthetic profile.
The NLTE corrections provided in this paper are applied to our LTE analysis
(see below).
\inlinecite{takeda96} select eight weak K lines, adding  to the list of
\inlinecite{zhang06} the lines at 404.7, 533.9, and 583.1\,nm, while they 
ignore the 1243.2\,nm line, which is included by \inlinecite{zhang06}.
According to \inlinecite{ivanova00}
NLTE corrections are small for the 693.8\,nm line, important for the 
1252.2, 1243.2, and 1176.9\,nm lines, and very important for the 769.8\,nm 
line.

Our present analysis is based on the six \ion{K}{i} lines listed in 
Table\,\ref{tbl:kilines}.
By means of the IRAF task {\tt splot} [\url{http://iraf.noao.edu/}], we have measured the equivalent
widths in the observed solar atlases. Interpolating in the curve-of-growth, 
we determine the Potassium abundance in the solar photosphere from both 3D 
and 1D model atmospheres. NLTE corrections were taken from 
\inlinecite{zhang06} where available.
The NLTE K abundance derived from the \xx\ model 
(with $\xi_{\rm mic}=1.0$\,\kms), $A$(K)$=5.06\pm 0.04$, 
is in good agreement with the value of 
$A$(K)=$5.12\pm 0.03$ obtained by \inlinecite{zhang06} by using the 
Kitt Peak integrated disc atlas, and excluding the uncertain 766.4\,nm 
line. The Potassium abundance given by \inlinecite{lodders09} is 
$A$(K)$=5.12\pm 0.03$, while \inlinecite{asplund09} recommend 
$A$(K)$=5.03\pm 0.09$.

%
 \begin{table}
 \caption{Atomic data and derived abundances for our selection 
          of \ion{K}{i} lines}
 \label{tbl:kilines}
 \begin{tabular}{rcccccc}     
 \hline
$\lambda$ & $\chi$ & \loggf & \multicolumn{4}{c}{$A$(K)}\\
          &       &        & 3D-LTE & \mD-LTE & \xx -LTE & 3D-NLTE\\
nm        & eV            & \\
 \hline
    580.1749 & 1.617 &  2.20 & 5.263 &  5.279 &  5.225 & 5.203 \\ 
    693.8763 & 1.617 &  4.00 & 4.998 &  5.017 &  4.962 & 4.929 \\ 
    769.8974 & 0.000 &170.00 & 5.434 &  5.458 &  5.353 & 5.144 \\ 
   1176.9689 & 0.000 & 50.00 & 5.329 &  5.359 &  5.288 & 5.199 \\ 
   1243.2273 & 1.617 & 56.00 & 5.279 &  5.308 &  5.237 & 5.129 \\ 
   1252.2134 & 1.610 & 85.00 & 5.275 &  5.304 &  5.222 & 5.055 \\ 
\\
   $\left\langle{\rm A(K)}\right\rangle$ & & & 5.263 & 5.296 & 5.223 & 5.110 \\
 \hline
 \end{tabular}
 \\
 \end{table}
 
Our adopted value is $A$(K)$_{\rm 3D-NLTE}=5.11\pm 0.09$, which is
obtained by applying the NLTE corrections of \inlinecite{zhang06} 
to the results from the 3D-LTE analysis for the integrated disc spectrum.
The 3D-LTE results from disc centre and integrated disc agree very
closely, the latter being 0.03\,dex higher. This tiny difference is
probably due to different NLTE corrections.

\subsection{Iron}

Iron is a complex atom with a very 
rich spectrum of atomic lines, especially from the neutral and singly ionised 
stage. Many scientists devoted time to the determination of the Iron abundance
in the solar photosphere, both from \ion{Fe}{i} (which accounts for most of the
lines in the solar spectrum) and from \ion{Fe}{ii} (which is the dominant 
ionisation stage of Iron in the photosphere).

The Iron abundance was continuously decreasing from the value given by 
\inlinecite{russell29} until the 1960s. The correction of the transition
probabilities produced an increasing solar abundance by about one order of
magnitude until the 1990s. Since then, $A$(Fe) has experienced a smooth decrease
(see Figure\,1 in \inlinecite{grevesse99}).
We recall some important developments in this field:
Corliss and his group investigated the Iron abundance,
with their transition probability \cite{corliss66,corliss68};
Bridges and coworkers applied their new $gf$-scale to the abundance
determination \cite{bridges70,bridges74}.
Between 1980--1990, there was a long debate between Holweger and the Blackwell 
group, advocating a low ($A$(Fe)$=7.50$) and a high ($A$(Fe)$=7.63$) Iron 
abundance, respectively \cite{biemont91,black1,holweger95,black2,kostik96}.
\inlinecite{kostik96} pointed out the necessity to perform a 3D-NLTE
analysis,  which was impossible at the time for the lack of computer 
power and atomic-physics data.

%
 \begin{table}
 \caption{Atomic data and equivalent widths for our selection 
          of 15 \ion{Fe}{ii} lines}
 \label{tbl:feiilines0}
 \begin{tabular}{cccccccc} 
 \hline
$\lambda$ & $\chi$ & \multicolumn{2}{c}{\loggf} & \multicolumn{2}{c}{EW [pm]} \\
nm        & eV            & H & MB & I & F \\
 \hline
   457.6340  &  2.84 & $-2.94$ & $-2.95$ & 6.90 & 6.90\\
   462.0521  &  2.83 & $-3.21$ & $-3.21$ & 5.60 & 5.50\\
   465.6981  &  2.89 & $-3.59$ & $-3.60$ & 3.60 & 3.50\\
   523.4625  &  3.22 & $-2.23$ & $-2.18$ & 8.80 & 8.60\\
   526.4812  &  3.23 & $-3.25$ & $-3.13$ & 4.70 & 4.70\\
   541.4073  &  3.22 & $-3.50$ & $-3.58$ & 2.90 & 2.80\\
   552.5125  &  3.27 & $-3.95$ & $-3.97$ & 1.35 & 1.25\\
   562.7497  &  3.39 & $-4.10$ & $-4.10$ & 0.82 & 0.84\\
   643.2680  &  2.89 & $-3.50$ & $-3.57$ & 4.40 & 4.35\\
   651.6080  &  2.89 & $-3.38$ & $-3.31$ & 5.80 & 5.80\\
   722.2394  &  3.89 & $-3.36$ & $-3.26$ & 1.95 & 1.92\\
   722.4487  &  3.89 & $-3.28$ & $-3.20$ & 2.30 & 2.03\\
   744.9335  &  3.89 & $-3.09$ & $-3.27$ & 1.75 & 1.80\\
   751.5832  &  3.90 & $-3.44$ & $-3.39$ & 1.60 & 1.50\\
   771.1724  &  3.90 & $-2.47$ & $-2.50$ & 5.30 & 5.12\\
\\
 \hline
 \end{tabular}
\\
H: \inlinecite{hannaford92}\\
MB: \inlinecite{melendez09}
 \end{table}

To investigate the Iron abundance in the solar photosphere, we selected
15 \ion{Fe}{ii} lines
and measured their equivalent widths with the IRAF task \url{splot}
(see Table\,\ref{tbl:feiilines}).
We presume that transitions from the dominant ionisation stage of Iron, 
\ion{Fe}{ii}, are close to LTE condition 
\cite{gehrenI}. 
This is not the case for \ion{Fe}{i}, for which NLTE-effects 
are important, at the level of $\approx 0.1$\,dex, as shown
by \citeauthor{gehrenI} (\citeyear{gehrenI,gehrenII}) using 
hydrostatic 1D model atmospheres, and by \inlinecite{ST01} using a 
single snapshot of a 3D hydrodynamical simulation.

For the disc centre and integrated disc spectra, the 3D abundance we 
find is $A$(Fe)$=7.525\pm 0.057$ and $A$(Fe)$=7.512\pm 0.062$, respectively,
when using the \loggf\ from \inlinecite{melendez09}.
When we use the \loggf\ from \inlinecite{hannaford92}, the 
line-to-line scatter increases and the abundances become
$A$(Fe)$=7.530\pm 0.110$ and $A$(Fe)$=7.516\pm 0.109$, respectively.
If we remove the 744.9\,nm line, which shows an exceptionally low
$A$(Fe) value, we slightly reduce the $\sigma$: 
$A$(Fe)$=7.550\pm 0.082$ and $A$(Fe)$=7.532\pm 0.093$, respectively.
Thirteen of our selected \ion{Fe}{ii} lines are in common
with those for which \inlinecite{Schnabel} 
have measured oscillator strengths. Using their set
of  \loggf\, values we obtain, for these 13 lines,
$A$(Fe)$_{\rm 3D} = 7.50 \pm 0.11$ and 
$A$(Fe)$_{\rm 3D} = 7.49 \pm 0.11$ for the 
for the disc centre and integrated disc spectra, respectively. 
We conclude that the different sets of oscillator strengths 
provide highly consistent Iron abundances. However,
the line-to-line scatter is considerably smaller when adopting 
the \inlinecite{melendez09} \loggf\, values. Therefore,
considering both disc centre and integrated disc spectra, 
we recommend $A$(Fe)$=7.52\pm 0.06$ for the photospheric 
solar Iron abundance.
As a consequence of the problem in representing the
turbulent motions on small scales in the hydrodynamical simulations, 
described in \cite{mst09}, there is a small slope of abundance
with equivalent width of the line, which amounts to
0.02 dex/pm.

The abundances derived from the 1D models depend on the choice of the 
microturbulence parameter.
When using the standard microturbulence value of 
$\xi_{\rm mic}=1.0$\,\kms, and 
the \loggf\, values of \inlinecite{melendez09}, the abundances are
$A$(Fe)$_{\mD}=7.45\pm 0.03$, $A$(Fe)$_{\rm HM}=7.48\pm 0.03$ 
for disc centre, and 
$A$(Fe)$_{\mD}=7.48\pm 0.05$, $A$(Fe)$_{\rm HM}=7.50\pm 0.05$ 
for integrated disc.
If we take instead the lower microturbulence values predicted by 
the 3D model, $\xi_{\rm mic}\approx 0.7$\,\kms\ for disc centre,
and $\xi_{\rm mic}\approx 0.9$\,\kms\ for the integrated disc
\cite{mst09}, the resulting Iron abundance is slightly higher by 
$\approx 0.04$ and $0.02$\, dex, respectively.

%
 \begin{table}
 \caption{Atomic data and derived abundances for our selection 
          of 15 \ion{Fe}{ii} lines}
 \label{tbl:feiilines}
 \begin{tabular}{cccccccc} 
 \hline
$\lambda$ & $\chi$ & \multicolumn{2}{c}{\loggf} & 
\multicolumn{4}{c}{$A$(Fe) (H$\,|\,$MB) } \\
nm        & eV            & H & MB & 3D\,(I) & 3D\,(F) & 
$\langle$3D$\rangle$\,(I) & $\langle$3D$\rangle$\,(F) \\
 \hline
   457.6340  &  2.84 & $-2.94$ & $-2.95$ & $7.63\,|\,7.64$ & $7.60\,|\,7.61$ & $7.49\,|\,7.50$ & $7.56\,|\,7.57$ \\
   462.0521  &  2.83 & $-3.21$ & $-3.21$ & $7.54\,|\,7.54$ & $7.48\,|\,7.48$ & $7.43\,|\,7.43$ & $7.45\,|\,7.45$ \\
   465.6981  &  2.89 & $-3.59$ & $-3.60$ & $7.45\,|\,7.46$ & $7.40\,|\,7.41$ & $7.39\,|\,7.40$ & $7.39\,|\,7.40$ \\
   523.4625  &  3.22 & $-2.23$ & $-2.18$ & $7.61\,|\,7.56$ & $7.58\,|\,7.53$ & $7.45\,|\,7.40$ & $7.51\,|\,7.46$ \\
   526.4812  &  3.23 & $-3.25$ & $-3.13$ & $7.67\,|\,7.55$ & $7.68\,|\,7.56$ & $7.58\,|\,7.46$ & $7.64\,|\,7.52$ \\
   541.4073  &  3.22 & $-3.50$ & $-3.58$ & $7.45\,|\,7.53$ & $7.42\,|\,7.50$ & $7.40\,|\,7.48$ & $7.40\,|\,7.48$ \\
   552.5125  &  3.27 & $-3.95$ & $-3.97$ & $7.45\,|\,7.47$ & $7.40\,|\,7.42$ & $7.42\,|\,7.44$ & $7.39\,|\,7.41$ \\
   562.7497  &  3.39 & $-4.10$ & $-4.10$ & $7.44\,|\,7.44$ & $7.45\,|\,7.45$ & $7.42\,|\,7.42$ & $7.44\,|\,7.44$ \\
   643.2680  &  2.89 & $-3.50$ & $-3.57$ & $7.47\,|\,7.54$ & $7.47\,|\,7.54$ & $7.39\,|\,7.46$ & $7.44\,|\,7.51$ \\
   651.6080  &  2.89 & $-3.38$ & $-3.31$ & $7.67\,|\,7.60$ & $7.69\,|\,7.62$ & $7.56\,|\,7.49$ & $7.64\,|\,7.57$ \\
   722.2394  &  3.89 & $-3.36$ & $-3.26$ & $7.59\,|\,7.49$ & $7.61\,|\,7.51$ & $7.55\,|\,7.45$ & $7.59\,|\,7.49$ \\
   722.4487  &  3.89 & $-3.28$ & $-3.20$ & $7.62\,|\,7.54$ & $7.57\,|\,7.49$ & $7.57\,|\,7.49$ & $7.54\,|\,7.46$ \\
   744.9335  &  3.89 & $-3.09$ & $-3.27$ & $7.25\,|\,7.43$ & $7.29\,|\,7.47$ & $7.21\,|\,7.39$ & $7.27\,|\,7.45$ \\
   751.5832  &  3.90 & $-3.44$ & $-3.39$ & $7.56\,|\,7.51$ & $7.54\,|\,7.49$ & $7.52\,|\,7.47$ & $7.52\,|\,7.47$ \\
   771.1724  &  3.90 & $-2.47$ & $-2.50$ & $7.54\,|\,7.57$ & $7.56\,|\,7.59$ & $7.43\,|\,7.46$ & $7.50\,|\,7.53$ \\
\\
 $\left\langle{\rm A(Fe)}\right\rangle$ & & & &  $7.53\,|\,7.53$ & $7.52\,|\,7.51$ & $7.46\,|\,7.45$ & $7.49\,|\,7.48$ \\
 \hline
 \end{tabular}
\\
H: \inlinecite{hannaford92}\\
MB: \inlinecite{melendez09}
 \end{table}

From our analysis, the recommended solar Iron abundance is 
A(Fe)\,=\,$7.52\pm 0.06$.

\subsection{Europium}

Basing our 3D LTE analysis \cite{mucciarelli08} 
on the five \ion{Eu}{ii} lines from \inlinecite{lawler01},
we obtain $A$(Eu)$=0.52\pm 0.02$. The isotopic ratio of 
$^{151}$Eu/($^{151}$Eu\,+\,$^{153}$Eu) derived from
disc centre and integrated disc spectra is
$(49\pm 2.3)$\% and $(50\pm 2.3)$\%, respectively.
We are in perfect agreement with the results of \inlinecite{lodders09}.

\subsection{Hafnium}

For the abundance determination of Hafnium \cite{thhf}, 
we considered four \ion{Hf}{ii}
lines, suggested by \inlinecite{lawler07}.
The lines are weak and blended, making the abundance determination a
difficult task.
Our abundance determination, $A$(Hf)$=0.87\pm 0.04$, is in close agreement
with the photospheric value of \inlinecite{lodders09}, but larger than
their meteoritic value of $0.73\pm 0.02$.

\subsection{Osmium}

Osmium and Iridium are among the heaviest stable elements and as such
are reliable reference elements for measuring the decay of the 
radioactive nuclei $^{238}$U and $^{232}$Th. 
Therefore Os is important in radioactive cosmochronology.

The determination of the solar Os abundance is not an easy task: the suitable 
\ion{Os}{i} lines lie mostly in the crowded violet and near UV range, and the 
transition probabilities have been revised several times in the last decades.
Improved centre-of-gravity wavelength are given by \inlinecite{ivarsson03}, 
who recall also that Os  has seven stable isotopes; the even isotopes 
which account for more than 80\% of all Osmium have no hyperfine structure.
However, according to the analysis of \inlinecite{quinet06}, the isotopic splitting and
the hyperfine structure of the odd isotopes have no influence on the 
Os abundance determination.

\inlinecite{ivarsson03} determined a new set of \loggf\ for 18 \ion{Os}{i} 
lines and compared them with previous values, finding a good agreement of 
their experimental results with previous determinations for the three strongest 
lines (290.9, 305.8, and 330.1\,nm). However, a substantial disagreement
is found for other lines with offset and dispersion increasing for the weaker
lines. The comparison with theoretical results shows an even larger scatter.
\inlinecite{ivarsson03} did not study the solar Os abundance, but applied 
their new \loggf\, values to the analysis of the metal poor giant 
CS 31082-001, previously analysed by \inlinecite{hill02}.

The solar abundance of Os has been determined by \inlinecite{jacoby76}
from three \ion{Os}{i} lines (305.8706, 330.1579, and 442.0460\,nm),
using \loggf\ from \inlinecite{CB}, to obtain $A$(Os)=0.70. Owing to the
large errors often present in the \inlinecite{CB} data, a new
determination of the lifetimes of nine transitions of Osmium, including the 
three lines used by \inlinecite{jacoby76}, has been done by 
\inlinecite{kwiatkowski84}. These new \loggf\ are systematically lower 
than those by \inlinecite{CB} (by about 0.5 dex) leading to 
$A$(Os)$=1.45\pm 0.10$ from the equivalent widths of nine lines measured 
in the solar atlas of \inlinecite{delbouille}, and using the HM model. 
This abundance agrees within one $\sigma$ with the Os abundance in 
meteorites of $A$(Os)$=1.37\pm 0.03$ \cite{lodders09}.

The most recent transition probabilities and solar Osmium abundance
are due to \inlinecite{quinet06}, who also give all of the possible
contaminations of each \ion{Os}{i} line in the Sun or in metal-poor
stars.  New \loggf\ are determined by these authors for 11 \ion{Os}{i}
lines and the difference with respect to the previous results is less
than 0.1~dex for all but the two lines at 397.2\,nm and 442.0\,nm.
Their newly determined \loggf\, value for the 
442.0\,nm line differs by 0.33 from the previous value; 
the new \loggf\, value of $-1.20$  produces an Os abundance in better
agreement with that from other \ion{Os}{i} lines in the star CS\,31082-001.
They determine the Os abundance using SYNTHE \cite{kurucz93b} 
with HM and MARCS models, but do not specify which observational data are used.
They discarded not only the heavily blended lines, but also the weakest
ones (EW$<0.2$\,pm) because they are faint, sensitive to blends, and more 
affected by uncertainties in the \loggf\, values.
In the end, \inlinecite{quinet06} retained only three lines (326.7945, 330.1559, 
and 442.0468\,nm) that give the abundances $A$(Os)= 1.40, 1.30, and 
1.15, respectively, with the HM model, and $A$(Os)= 1.30, 1.25, and 
1.10 with a MARCS model. 
Their final Os abundance is $A$(OS)$=1.25 \pm 0.11$.

We adopted the equivalent widths of the \ion{Os}{i} lines from
\inlinecite{kwiatkowski84} and derived the abundances interpolating in
theoretical curves-of-growth (see Table\,\ref{tbl:osilines}). 
We adopted the partition function from \inlinecite{Allen}.
If we
weight the abundances from the individual lines as in their analysis, we
obtain an abundance of $A$(Os)$_{\rm 3D}=1.36\pm 0.19$. The straight
average gives $A$(Os)$_{\rm 3D}=1.35\pm 0.21$. The \loggf\, values that we use are
from \inlinecite{quinet06} for all lines except for the 409\,nm,
442\,nm, and 463\,nm lines for which we use the \loggf\, values from
\inlinecite{kwiatkowski84}. If we restrict our selection to the three
lines considered most reliable by \inlinecite{quinet06} 
(330.1\,nm, 409.1\,nm, and 455.0\,nm) we obtain a very similar value 
for the average ($A$(Os)=1.37) with an improved but still very large
line-to-line scatter ($\sigma$=0.16). For this reason we recommend
the result from the complete sample of lines.

 \begin{table}
 \caption{\ion{Os}{i} lines used in the analysis with the abundance obtained
 from the EWs of disk-centre using 3D, \mD, and \xx\ model atmospheres.
 The microturbulence was set to 1.0\,\kms\ for the 1D models.}
 \label{tbl:osilines}
 \begin{tabular}{cccccccc}
 \hline
 $\lambda$ & EW & $\chi$ & \loggf\ &  \multicolumn{3}{c}{$A$(Os)} & ${\rm Weight}^Q$\\
    nm     & pm & eV    &         & 3D & \mD & \xx & \\
 \hline
 305.8660 & 1.03  & 0.00  & --0.41$^Q$ & 1.070  & 1.069 &  1.020 & 1 \\
 326.7945 & 0.39  & 0.00  & --1.09$^Q$ & 1.211  & 1.228 &  1.177 & 1 \\
 330.1559 & 0.72  & 0.00  & --0.74$^Q$ & 1.157  & 1.165 &  1.114 & 3 \\    
 375.2524 & 0.50  & 0.34  & --0.98$^Q$ & 1.442  & 1.451 &  1.406 & 1 \\
 397.7231 & 0.07  & 0.64  & --1.94$^Q$ & 1.777  & 1.789 &  1.744 & 1 \\
 409.1817 & 0.06  & 0.76  & --1.66$^K$ & 1.533  & 1.544 &  1.499 & 3 \\
 442.0468 & 0.23  & 0.00  & --1.53$^K$ & 1.195  & 1.218 &  1.165 & 1 \\
 455.0410 & 0.04  & 1.84  & --0.71$^K$ & 1.415  & 1.412 &  1.371 & 3 \\
 463.1828 & 0.01  & 1.89  & --1.19$^K$ & 1.332  & 1.329 &  1.288 & 1 \\
\\
 $\left\langle{\rm A(Os)}\right\rangle$ & & & & 1.348 & 1.356 & 1.309 &  \\
 \hline
 \end{tabular}                                              
 \\
 Weight according to \inlinecite{quinet06}\\
 $^Q$ \inlinecite{quinet06}\\
 $^K$ \inlinecite{kwiatkowski84}\\
 \\
 \end{table}

\begin{figure} 
 \centerline{\includegraphics[width=0.8\textwidth,clip=]{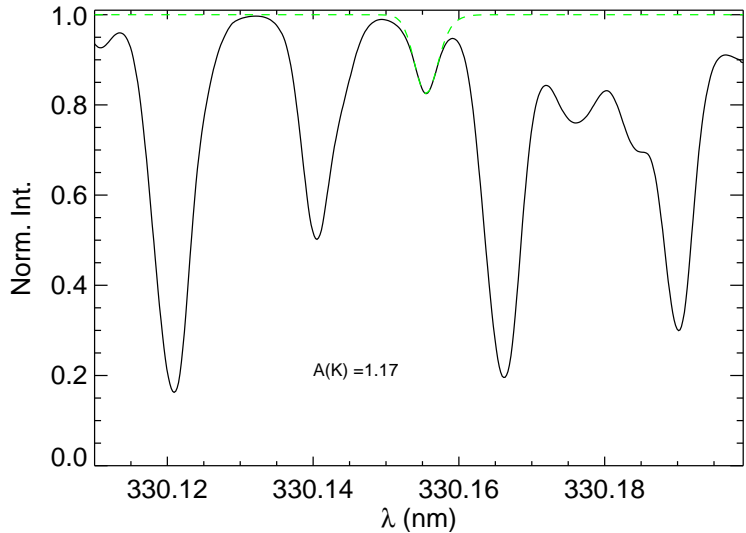}}
 \caption{The 3D synthetic profile the \ion{Os}{i} line at 330.1\,nm 
  with $A$(Os)=1.17 (green dashed line) is over-imposed on the disc centre 
  solar atlas of \protect\inlinecite{delbouille} (black solid line).}
 \label{fig:os3301}
 \end{figure}

 As an additional check, we fitted the profile of the line at 330.1\,nm,
 which is the cleanest of our lines, with the \mD+SYNTHE synthetic spectrum, 
 allowing us to take into account all of the blending lines present in the 
 region. The abundance we find from this fit is in very good agreement with 
 the 1D result derived from the equivalent width. As a further illustration,
 we show in Figure\,\ref{fig:os3301} the observed profile of this line 
 over-plotted with the 3D synthetic profile. Without the blending lines 
 included, it is difficult to place the continuum in this case.

%

\subsection{Thorium}

The only line of Thorium in the solar spectrum that can 
be used for abundance determination is the \ion{Th}{ii} resonance
line at 401.9\,nm \cite{thhf}. 
This weak absorption lies on the red wing 
of a stronger Fe-Ni feature and is blended with weaker lines.
We performed a line profile fitting analysis with a 3D
synthetic profile, in order to take into account the convection-induced
asymmetry, of the Fe-Ni blend.
Giving higher weight to the results from the disc centre data, which resolve
the line profile better, we obtained $A$(Th)$=0.08\pm 0.03$ \cite{thhf}.
This result is in perfect agreement with the meteoritic value
\cite{lodders09}.

\section{Conclusions}

The abundances of the solar photosphere,
based on our \cobold\ solar model atmosphere,
are reported in Table\,\ref{tbl:abbo}.
We find that 3D corrections are small in general, and not responsible 
for the systematic lowering of the solar abundances over the past years. 
With the abundances given in Table\,\ref{tbl:abbo}, and the abundances 
from \inlinecite{lodders09}, photospheric where available, 
for the elements that we did not analyse, we obtain 
a solar metallicity of $Z=0.0153$, and $Z/X=0.0209$.
This value can be compared to 
$Z=0.0141$, and $Z/X=0.0191$ of \inlinecite{lodders09} and 
$Z=0.0134$, and $Z/X=0.0181$ of \inlinecite{asplund09}.

Our spectroscopic abundances of O and Fe are in agreement, 
within the indicated mutual errors, with what is derived from 
helioseismic constraints: $A$(O)$=8.86\pm 0.045$, 
$A$(Fe)$=7.50\pm 0.048$ \cite{DP06}.
There is also a good agreement for the key elements C, N and O,  
with the most likely \emph{spectroscopic} abundances recommended by 
\inlinecite{PD09}: $A$(C)$=8.44\pm 0.06$, $A$(N)$=7.96\pm 0.10$, 
$A$(O)$=8.75\pm 0.08$.

We conclude that our derived solar metallicity goes in the direction
of reconciling atmospheric abundances with helioseismic data.
According to \citeauthor{turcotte02} (\citeyear{turcotte02}), 
the abundance of metals in the solar 
photosphere may have decreased by up to 0.04\,dex by diffusion
during the solar lifetime. This would improve the correspondence
between spectroscopic and helioseismic results even further.

\begin{table}
\caption{The recommended abundances of the solar photosphere for 
          the elements considered in this work.}
\label{tbl:abbo}
\begin{tabular}{llcrl}
\hline
Element & Ion.  & Abundance & N & Reference \\
        & state &           & lines\\
\hline
Li & {\sl I}  & $1.03\pm 0.03$ & 1  & \\
C  & {\sl I}  & $8.50\pm 0.06$ & 45 & \inlinecite{carbon}\\
N  & {\sl I}  & $7.86\pm 0.12$ & 12 & \inlinecite{azoto}\\
O  & {\sl I}  & $8.76\pm 0.07$ & 10 & \inlinecite{oxy}\\
P  & {\sl I}  & $5.46\pm 0.04$ & 5  & \inlinecite{phosphorus}\\
S  & {\sl I}  & $7.16\pm 0.05$ & 7  & \inlinecite{zolfito},\inlinecite{zolfo}\\
K  & {\sl I}  & $5.11\pm 0.09$ & 6  & \\
Fe & {\sl II} & $7.52\pm 0.06$ & 15 & \\
Eu & {\sl II} & $0.52\pm 0.03$ & 5  & \inlinecite{mucciarelli08}\\
Hf & {\sl II} & $0.87\pm 0.04$ & 4  & \inlinecite{thhf}\\
Os & {\sl I}  & $1.36\pm 0.19$ & 9  & \\
Th & {\sl II} & $0.08\pm 0.03$ & 1  & \inlinecite{thhf}\\
\hline                                                     
\end{tabular}                                              
\\
\end{table}

%


 \begin{acks}
We thank Rosanna Faraggiana for her constant support and precious advise.
We are grateful to Sergei Andrievsky and Sergei Korotin for NLTE computations 
and useful discussions.
We acknowledge support from EU contract MEXT-CT-2004-014265 (CIFIST).
 \end{acks}

%
%
%
%
%
%

\end{article} 
\end{document}